\documentclass[preprint,superscriptaddress,preprintnumbers,showpacs]{elsart}
\usepackage{amssymb}
\usepackage{graphicx}
\usepackage{dcolumn}
\usepackage{bm}

\journal{Physics Letter B}

\begin{document}

\begin{frontmatter}

\title{A new simple form of quark mixing matrix}

\author[1]{Nan Qin},
\author[1,2]{Bo-Qiang Ma\corauthref{*}} \corauth[*]{Corresponding author at: School of Physics,
Peking University, Beijing 100871, China.}\ead{mabq@pku.edu.cn}
\address[1]{School of Physics and State Key Laboratory of Nuclear Physics and
Technology, Peking University, Beijing 100871, China}
\address[2]{Center for High Energy
Physics, Peking University, Beijing 100871, China}

\begin{abstract}
Although different parametrizations of quark mixing matrix are
mathematically equivalent, the consequences of experimental analysis
may be distinct. Based on the triminimal expansion of
Kobayashi-Maskawa matrix around the unit matrix, we propose a new
simple parametrization. Compared with the Wolfenstein
parametrization, we find that the new form is not only consistent
with the original one in the hierarchical structure, but also more
convenient for numerical analysis and measurement of the
CP-violating phase. By discussing the relation between our new form
and the unitarity boomerang, we point out that along with the
unitarity boomerang, this new parametrization is useful in hunting
for new physics.
\end{abstract}


\end{frontmatter}

\newpage

The mixing of quarks is one of the fundamental problems in particle
physics. However, its origin is still unclear yet and the mixing is
currently described phenomenologically by the mixing matrix, i.e.,
the Cabibbo\cite{cabibbo}-Kobayashi-Maskawa\cite{km}(CKM) matrix
\begin{eqnarray}
V_{\rm CKM}=\left(
  \begin{array}{ccc}
    V_{ud} & V_{us} & V_{ub} \\
    V_{cd} & V_{cs} & V_{cb} \\
    V_{td} & V_{ts} & V_{tb} \\
  \end{array}\right).
\end{eqnarray}
The parametrization proposed by Chau and Keung (CK)~\cite{ck,pdg} is
the most popular way of parameterizing the matrix. Using three
mixing angles and one CP-violating phase, it provides a clear
understanding of the mixing. However, some recent
works~\cite{he,lishiwen} reveal that the parameters in the CK
parametrization are inconvenient when dealing with the unitarity
boomerang (UB). A unitarity boomerang is formed using two unitarity
triangles~\cite{ut} with a common inner angle, thus contains all
four independent parameters in the mixing matrix, and is a powerful
tool of hunting for new physics beyond the Standard
Model~\cite{boom}. Instead of the CK form, Frampton and He
proposed~\cite{he} that the original Kobayashi-Maskawa
(KM)~\cite{km} matrix is kept as the standard parametrization, which
is given by
\begin{eqnarray}
V&=&\left(
  \begin{array}{ccc}
    1 & 0   & 0    \\
    0 & c_2 & -s_2 \\
    0 & s_2 & c_2  \\
  \end{array}
\right)\left(
  \begin{array}{ccc}
    c_1 & -s_1 & 0 \\
    s_1 & c_1  & 0 \\
    0   & 0    & e^{i\delta} \\
  \end{array}
\right)\left(
  \begin{array}{ccc}
    1 & 0   & 0    \\
    0 & c_3 & s_3  \\
    0 & s_3 & -c_3 \\
  \end{array}
\right)\nonumber\\
&=&\left(
\begin{array}{ccc}
c_1     & -s_1c_3                     & -s_1s_3           \\
s_1c_2  & c_1c_2c_3-s_2s_3e^{i\delta} & c_1c_2s_3+s_2c_3e^{i\delta} \\
s_1s_2  & c_1s_2c_3+c_2s_3e^{i\delta} & c_1s_2s_3-c_2c_3e^{i\delta}
\end{array}
\right)\;.\label{KM}
\end{eqnarray}
Here $s_i=\sin\theta_i$, $c_i=\cos\theta_i$ $(i=1,2,3)$, and
$\theta_1$, $\theta_2$, $\theta_3$ are Euler angles describing the
rotation among different generations, $\delta$ is the CP-violating
phase in the KM parametrization.

Although different parametrizations of quark mixing matrix are
mathematically equivalent, the consequences of experimental analysis
may be distinct. The magnitudes of the elements $V_{ij}$ are
physical quantities which do not depend on parametrization. However,
the CP-violating phase does. As a result, the understanding of the
origin of CP violation is associated with the parametrization. For
example, the prediction based on the maximal CP-violation
hypothesis~\cite{maximalviolation} is related with the
parametrization or in other words, phase convention. As discussed in
Ref.~\cite{koide}, only with the original KM parametrization and the
Fritzsch-Xing~\cite{xing} parametrization, one can get successful
predictions on the unitarity triangle~\cite{ut} from the maximal
CP-violation hypothesis. Therefore the original KM matrix is
convenient for studying both the maximal CP-violation and unitarity
boomerangs, so that a study about it is necessary.

With the data on the magnitudes of the mixing matrix elements~\cite{pdg}
\begin{eqnarray} \left(
  \begin{array}{ccc}
    0.97419\pm0.00022             & 0.2257\pm0.0010    & 0.00359\pm0.00016               \\
    0.2256\pm0.0010               & 0.97334\pm0.00023  & 0.0415^{+0.0010}_{-0.0011}      \\
    0.00874^{+0.00026}_{-0.00037} & 0.0407\pm0.0010    & 0.999133^{+0.000044}_{-0.000043}
  \end{array} \right)\;,\label{ckmdata}
\end{eqnarray}
one can easily get the ranges of the parameters in the KM parametrization
\begin{eqnarray}
\theta_1=0.228\pm0.001,\quad \theta_2=0.039_{-0.002}^{+0.001},\quad \theta_3=0.016\pm0.001\;.\label{kmdata}
\end{eqnarray}
When studying mixing, it is useful to parameterize the matrix
according to the hierarchical structure of the mixing to reveal more
physical information about the underlying theory. A good choice is
the idea of  triminimal
parametrization~\cite{triminimalneutrino,triminimalquark,triminimalquark2}
with an approximation as the basis matrix to the lowest order. That
is to express a mixing angle in the mixing matrix as the sum of a
zeroth order angle $\theta^0$ and a small perturbation angle
$\epsilon$ with
\begin{eqnarray}
\theta_1=\theta_1^0+\epsilon_1,\quad\theta_2=\theta_2^0+\epsilon_2,\quad\theta_3=\theta_3^0+\epsilon_3.
\end{eqnarray}
With the deviations $\epsilon_i$, one can expand the mixing matrix in powers of $\epsilon_i$ while different choices
of $\theta_i^0$ lead to different basis. The general expansion of KM matrix is presented in Appendix A.
Since Eq.(\ref{ckmdata}) is very close to the unit matrix, it is a
good approximation to let
\begin{eqnarray}
\epsilon_1=\theta_1,\quad\epsilon_2=\theta_2,\quad\epsilon_3=\theta_3\;.
\end{eqnarray}
To make the lowest order be the unit matrix, we still need to adjust the phases of quarks with
\begin{eqnarray}
c\rightarrow ce^{i\pi},\quad s\rightarrow se^{i\pi},\quad b\rightarrow be^{i(\pi+\delta)}\;.\label{redefineq}
\end{eqnarray}
According to Eq.~(\ref{kmdata}), we have
$\epsilon_1^2\sim\epsilon_2\sim\epsilon_3$. Therefore, in order to
keep the magnitude consistency of the expansion, we display all
terms of $\mathcal {O}(\epsilon_1^3)$ in our parametrization with
\begin{eqnarray}
V=\left( \begin{array}{ccc}
1-\frac{\epsilon_1^2}{2}&\epsilon_1-\frac{\epsilon_1^3}{6}&e^{-i\delta}\epsilon_1\epsilon_3\\
\frac{\epsilon_1^3}{6}-\epsilon_1&1-\frac{\epsilon_1^2}{2}&\epsilon_2+e^{-i\delta}\epsilon_3\\
\epsilon_1\epsilon_2&-\epsilon_2-e^{i\delta}\epsilon_3&1
\end{array}\right)
+\mathcal {O}(\epsilon_1^4)\;.\label{unitexpand}
\end{eqnarray}
Comparing with the Wolfenstein parametrization~\cite{wolf}
\begin{eqnarray}
V&=&\left( \begin{array}{ccc}
1-\frac{1}{2}\lambda^2&\lambda&A\lambda^3(\rho-i\eta)\\
-\lambda&1-\frac{1}{2}\lambda^2&A\lambda^2\\
A\lambda^3(1-\rho-i\eta)&-A\lambda^2&1
\end{array}\right)
+\mathcal {O}(\lambda^4)\;,\label{wolf}
\end{eqnarray}
in which $\lambda=s_1$,
$A\lambda^2(\rho^2+\eta^2)^{\frac{1}{2}}=s_3$ and
$A\lambda^2[(1-\rho)^2+\eta^2]^{\frac{1}{2}}=s_2$,
Eq.~(\ref{unitexpand}) has the same hierarchical structure with the
Wolfenstein parametrization. We can check the magnitude consistency
by substituting these relations into Eq.~(\ref{unitexpand}) and only
focus on the modulus of each element in terms of all four
Wolfenstein parameters, which gives
\begin{eqnarray}
\left(\begin{array}{ccc}
1-\frac{1}{2}\lambda^2&\lambda&A\lambda^3(\rho^2+\eta^2)^{\frac{1}{2}}\\
\lambda&1-\frac{1}{2}\lambda^2&A\lambda^2(1-2\rho+2\rho^2+2\eta^2)^{\frac{1}{2}}\\
A\lambda^3((1-\rho)^2+\eta^2)^{\frac{1}{2}}&~A\lambda^2(1-2\rho+2\rho^2+2\eta^2)^{\frac{1}{2}}&1
\end{array}\right). ~~~~~
\label{moduli}
\end{eqnarray}
Here we take $\delta\approx90\textordmasculine$, which implies the
maximal CP violation. The only difference comes from $|V_{cb}|$ and
$|V_{ts|}$ with an extra coefficient. However, numerical calculation
gives $(1-2\rho+2\rho^2+2\eta^2)^{\frac{1}{2}}=1.0089\approx1$, so
that the hierarchical structure of the quark mixing is well
preserved in Eq.~(\ref{unitexpand}).

A natural idea is to find the relation between these two forms.
However, it is complicated in adjusting the phases by rephasing the
quark fields, as shown in Ref.~\cite{lishiwen}. This is because the
phase convention adopted by Eq.~(\ref{unitexpand}) is different from
Eq.~(\ref{wolf}). Actually, the Wolfenstein parametrization takes
the same phase convention with the standard CK form~\cite{ck,pdg},
which implies another choice of the phase $\delta$. Therefore one
has difficulty to arrive at the Wolfenstein parametrization from
triminimal parametrization of KM matrix. This is different from the
situation of triminimal parametrization of CK matrix, as shown in
Ref.~\cite{triminimalquark2}, where the Wolfenstein parametrization
can be understood as a simple form ``derived" from the CK matrix.

By keeping the original Wolfenstein parameter
$\lambda=\sin{\theta_1}\approx\epsilon_1-\frac{\epsilon_1^3}{6}$ and
the CP-violating phase $\delta$, and introducing two new parameters
with
\begin{eqnarray}
f\lambda^2=\sin{\theta_2}\approx\epsilon_2, \quad
h\lambda^2=\sin{\theta_3}\approx\epsilon_3\;,\label{fhrelation}
\end{eqnarray}
we obtain a new Wolfenstein-like parametrization through
substitution of them into Eq.~(\ref{unitexpand}), that is
\begin{eqnarray}
V=\left(\begin{array}{ccc}
1-\frac{\lambda^2}{2}&\lambda&e^{-i\delta}h\lambda^3\\
-\lambda&1-\frac{\lambda^2}{2}&(f+e^{-i\delta}h)\lambda^2\\
f\lambda^3&-(f+e^{i\delta}h)\lambda^2&1
\end{array}\right)
+\mathcal {O}(\lambda^4)\;.\label{newwolf}
\end{eqnarray}
This new simple form obviously preserves the unitarity of the matrix
to the third order of $\lambda$ and the hierarchical structure of
the quark mixing as we discussed above. The choice of two new
parameters is quite natural since $h\lambda^3=|V_{ub}|$ and
$f\lambda^3=|V_{td}|$, thus can directly be determined with
$\lambda=0.2257_{-0.0010}^{+0.0009}$~\cite{pdg} and
Eq.~(\ref{ckmdata}), which gives
\begin{eqnarray}
h=0.312^{+0.018}_{-0.014},\quad f=0.760^{+0.023}_{-0.032}\;.
\end{eqnarray}
Different from the original Wolfenstein form, in which the CP
violation is determined by two parameters,  i.e., $\rho$ and $\eta$,
there is only one phase $\delta$ independent of other parameters in
Eq.~(\ref{newwolf}). Another advantage of this new form is that
$V_{cb}$ and $V_{ts}$, with magnitudes of $\mathcal {O}(10^{-2})$,
contribute to the constraint of CP-violating phase $\delta$, while
in the original Wolfenstein form we need to consider $V_{ub}$ and
$V_{td}$, whose magnitudes being one order smaller than those of
$V_{cb}$ and $V_{ts}$ but with all four parameters involved, making
it inconvenient when doing experimental analysis. Therefore, from
this point of view, Eq.~(\ref{newwolf}) is more convenient than the
original Wolfenstein parametrization. Simple numerical calculation
of equation $|(f+e^{-i\delta}h)\lambda^2|=|V_{cb}|$ gives
\begin{eqnarray}
\delta\approx91.4\textordmasculine\;,\label{cpphase}
\end{eqnarray}
which means approximate maximal CP violation as we mentioned before.

\begin{figure}[!ht]
\begin{center}
\includegraphics[width=0.5\textwidth]{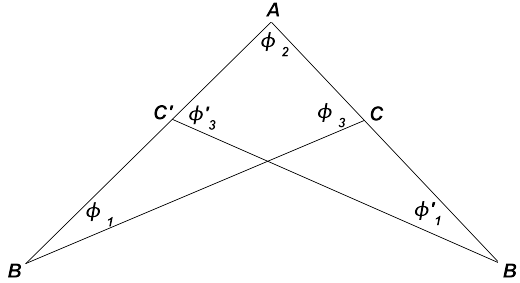}
\caption{The unitarity boomerang of quark mixing with the common
angle $\phi_2$. The sides are $AC=|V_{ud}V_{ub}^*|$,
$AC^{\prime}=|V_{ub}V_{tb}^*|$, $AB=|V_{td}V_{tb}^*|$,
$AB^{\prime}=|V_{ud}V_{td}^*|$, $BC=|V_{cd}V_{cb}^*|$,
$B^{\prime}C^{\prime}=|V_{us}V_{ts}^*|$. \label{fig1}}
\end{center}
\end{figure}

A useful and natural application of this new simple parametrization
is to study the unitarity boomerangs with it. The commonly used
unitarity boomerang is consisted by two unitarity triangles with the
same order of the three sides, say, $\lambda^3$, arising from
\begin{eqnarray}
V_{ud}V^*_{ub}+V_{cd}V^*_{cb}+V_{td}V_{tb}^*=0,\quad
V_{ud}V^*_{td}+V_{us}V^*_{ts}+V_{ub}V_{tb}^*=0.\label{uts}
\end{eqnarray}
Since the common angle of the two chosen unitarity boomerangs could
be determined by the CP-violating measurement
$J$~\cite{jarlskog,Wu:1985ea}, the CP-violating phase could then be
constrained. The Jarlskog parameter satisfies
\begin{eqnarray}
J&=&2|V_{td}V_{tb}^*||V_{ud}V_{ub}^*|\sin{\phi_2}\nonumber\\
&=&2|V_{ud}V_{td}^*||V_{ub}V_{tb}^*|{\sin{\phi_2}}'\nonumber
\end{eqnarray}
with $\phi_2=\phi'_2$ as the common angle of the unitarity boomerang
as illustrated in Fig.~\ref{fig1}.
Using
Eq.~(\ref{newwolf}), we easily parameterize the sides and angles of
the unitarity boomerang with
\begin{eqnarray}
AB&=&AB^{\prime}=f\lambda^3;\nonumber\\
AC&=&AC^{\prime}=h\lambda^3;\nonumber\\
BC&=&B^{\prime}C^{\prime}=\lambda^3(f^2+2fh\cos{\delta}+h^2)^{\frac{1}{2}};\nonumber\\
\phi_1&=&\phi_1'=\arctan{\frac{h\sin{\delta}}{f+h\cos{\delta}}};\nonumber\\
\phi_3&=&\phi_3'=\arctan{\frac{f\sin{\delta}}{h+f\cos{\delta}}};\nonumber\\
\phi_2&=&\phi_2'=\pi-\delta\;,\nonumber
\end{eqnarray}
showing that to the third order of $\lambda$, the two chosen
unitarity triangles are identical. Using the last one of these
equations, we can check the maximal CP-violation
hypothesis~\cite{koide} easily, and the experimental analysis
consistently gives
$\phi_2=(88_{-5}^{+6})\textordmasculine$~\cite{maximalcp}. High
order corrections to the boomerang bring about difference between
these two triangles (see Appendix B). To the lowest order, the
Jarlskog parameter is given by
\begin{eqnarray}
J=fh\lambda^6\sin{\delta}\;.\nonumber
\end{eqnarray}

We get simple relations between these two parametrizations, {\it
i.e.}, diagrammatical and matrix forms. This implies that the
parametrization Eq.~(\ref{newwolf}) is natural in discussing the
unitarity boomerangs of quark mixing. In Ref.~\cite{he} and
Ref.~\cite{boom}, Frampton and He pointed out that the unitarity
boomerang is very helpful in searching new physics since it contains
all the information about the mixing matrix and reflects the
precision attained by high-energy experiments. Thus deviations from
the expected unitarity boomerang may imply possibility for new
physics beyond the Standard Model. Therefore, if new physics
information show up in the unitarity boomerang analysis, we could
get corresponding signals in the parameters and consequently the
mixing matrix through the relations above. Then by studying how the
new physics modify the original matrix, we may get hints of
understanding the underlying theory.

Finally, we present a conclusion of this Letter. The new form of
quark mixing matrix Eq.~(\ref{newwolf}) is our main result. It
exhibits the hierarchical structure of the mixing, and is convenient
for numerical analysis, especially for constraint of the
CP-violating phase. Combined with the unitarity boomerang, it is
also helpful to study the presence of new physics. Therefore, we
humbly suggest it as a simple form corresponding to the KM matrix in
theoretical and experimental studies.

This work is partially supported by National Natural Science
Foundation of China (Nos.~10721063, 10975003, 11035003) and by the
Key Grant Project of Chinese Ministry of Education (No.~305001).

\appendix

{\bf{Appendix A: The general triminimal expansion of the KM matrix}}

We present here the general triminimal expansion of KM matrix. To
second order of $\epsilon_i$, the KM matrix is given by
\begin{eqnarray}
V&=&\left( \begin{array}{ccc} c_1^0 & -s_1^0 c_3^0 & -s_1^0 s_3^0
\\s_1^0c_2^0 & c_1^0 c_2^0 c_3^0-s_2^0 s_3^0 e^{i\delta} & c_1^0 c_2^0 s_3^0+s_2^0 c_3^0 e^{i\delta}
\\s_1^0s_2^0&c_1^0s_2^0c_3^0+c_2^0s_3^0e^{i\delta}&c_1^0s_2^0s_3^0-c_2^0c_3^0e^{i\delta}
\end{array}\right)
+\epsilon_1\left( \begin{array}{ccc} -s_1^0 & -c_1^0c_3^0
&-c_1^0s_3^0
\\ c_1^0c_2^0 & -c_2^0c_3^0s_1^0 & -c_2^0s_1^0s_3^0 \\
c_1^0s_2^0 & -c_3^0s_1^0s_2^0 &-s_1^0s_2^0s_3^0
\end{array}\right)\nonumber\\
&+&\epsilon_2\left(\begin{array}{ccc}
0&0&0\\
-s_1^0s_2^0 &-c_1^0c_3^0s_2^0-c_2^0s_3^0e^{i\delta} &-c_1^0s_2^0s_3^0+c_2^0c_3^0e^{i\delta}\\
c_2^0s_1^0
&c_1^0c_2^0c_3^0-s_2^0s_3^0e^{i\delta}&c_1^0c_2^0s_3^0+c_3^0s_2^0e^{i\delta}
\end{array}\right)\nonumber\\
&+&\epsilon_3\left( \begin{array}{ccc}
0&s_1^0s_3^0&-c_3^0s_1^0\\
0&-c_1^0c_2^0s_3^0-c_3^0s_2^0e^{i\delta}&c_1^0c_2^0c_3^0-s_2^0s_3^0e^{i\delta}\\
0&-c_1^0s_2^0s_3^0+c_2^0c_3^0e^{i\delta}&c_1^0c_3^0s_2^0+c_2^0s_3^0e^{i\delta}
\end{array}\right)
+\frac{1}{2}\epsilon_1^2\left(\begin{array}{ccc}
-c_1^0&c_3^0s_1^0&s_1^0s_3^0\\
-c_2^0s_1^0&-c_1^0c_2^0c_3^0&-c_1^0c_2^0s_3^0\\
-s_1^0s_2^0&-c_1^0c_3^0s_2^0&-c_1^0s_2^0s_3^0
\end{array}\right)\nonumber\\
&+&\frac{1}{2}\epsilon_2^2\left(\begin{array}{ccc}
0&0&0\\
-c_2^0s_1^0&-c_1^0c_2^0c_3^0+s_2^0s_3^0e^{i\delta}&-c_1^0c_2^0s_3^0-c_3^0s_2^0e^{i\delta}\\
-s_1^0s_2^0&-c_1^0c_3^0s_2^0-c_2^0s_3^0e^{i\delta}&-c_1^0s_2^0s_3^0+c_2^0c_3^0e^{i\delta}
\end{array}\right)\nonumber\\
&+&\frac{1}{2}\epsilon_3^2\left(\begin{array}{ccc}
0&c_3^0s_1^0&s_1^0s_3^0\\
0&-c_1^0c_2^0c_3^0+s_2^0s_3^0e^{i\delta}&-c_1^0c_2^0s_3^0-c_3^0s_2^0e^{i\delta}\\
0&-c_1^0c_3^0s_2^0-c_2^0s_3^0e^{i\delta}&-c_1^0s_2^0s_3^0+c_2^0c_3^0e^{i\delta}
\end{array}\right)
+\epsilon_1\epsilon_2\left( \begin{array}{ccc}
0&0&0\\
-c_1^0s_2^0&c_3^0s_1^0s_2^0&s_1^0s_2^0s_3^0\\
c_1^0c_2^0&-c_2^0c_3^0s_1^0&-c_2^0s_1^0s_3^0
\end{array}\right)\nonumber\\
&+&\epsilon_2\epsilon_3\left( \begin{array}{ccc}
0&0&0\\
0&c_1^0s_2^0s_3^0-c_2^0c_3^0e^{i\delta}&-c_1^0c_3^0s_2^0-c_2^0s_3^0e^{i\delta}\\
0&-c_1^0c_2^0s_3^0-c_3^0s_2^0e^{i\delta}&c_1^0c_2^0c_3^0-s_2^0s_3^0e^{i\delta}
\end{array}\right)
+\epsilon_1\epsilon_3\left(\begin{array}{ccc}
0&c_1^0s_3^0&-c_1^0c_3^0\\
0&c_2^0s_1^0s_3^0&-c_2^0c_3^0s_1^0\\
0&s_1^0s_2^0s_3^0&-c_3^0s_1^0s_2^0
\end{array}\right)\nonumber\\
&+&\mathcal {O}(\epsilon_i^3)\;,\label{general}\nonumber
\end{eqnarray}
where $s_i^0=\sin{\theta_i^0}$ and $c_i^0=\cos{\theta_i^0}$.

The Jarlskog parameter given by
\begin{eqnarray}
J={\rm Im}(V_{11}V_{22}V_{12}^*V_{21}^*)=s_1^2s_2s_3c_1c_2c_3\sin{\delta}\label{jarskog}\nonumber
\end{eqnarray}
is independent of phase convention, making it important when discussing CP violation. Expanding $J$ with $\epsilon_i$
to the second order gives
\begin{eqnarray}
J&=&J_0(1+\epsilon_1(3\cot{2\theta_1^0}+\csc{2\theta_1^0})+2\epsilon_2\cot{2\theta_2^0}+2\epsilon_3\cot{2\theta_3^0}
+\frac{1}{4}\epsilon_1^2(9\cos{2\theta_1^0}-5)\csc^2{\theta_1^0}\nonumber\\
&-&2\epsilon_2^2-2\epsilon_3^2+2\epsilon_1\epsilon_2(3\cos{2\theta_1^0}+1)\cot{2\theta_2^0}\csc{2\theta_1^0}
+4\epsilon_2\epsilon_3\cot{2\theta_2^0}\cot{2\theta_3^0}\nonumber\\
&+&2\epsilon_1\epsilon_3(3\cos{2\theta_1^0}+1)\cot{2\theta_3^0}\csc{2\theta_1^0})
+\mathcal {O}(\epsilon_i^3)\;,\label{generalj}\nonumber
\end{eqnarray}
in which $J_0=(s_1^0)^2s_2^0s_3^0c_1^0c_2^0c_3^0\sin{\delta}$.

{\bf{Appendix B: High order calculation of the boomerang}}

The leading order of the sides of the unitarity boomerang in
Fig.~\ref{fig1} are of $\mathcal {O}(\lambda^3)$ and the two
unitarity triangles are identical with each other. When high order
corrections are included, difference between the two triangles comes
out. We need to parameterize the CKM matrix to high order of
$\lambda$, here we expand it to $\mathcal {O}(\lambda^5)$,
\begin{eqnarray}
V=\left(\begin{array}{ccc}
1-\frac{\lambda^2}{2}-\frac{\lambda^4}{8}&\lambda-\frac{h^2\lambda^5}{2}&e^{-i\delta}h\lambda^3\\
-\lambda+\frac{f^2\lambda^5}{2}&1-\frac{\lambda^2}{2}-\frac{1}{8}(1+4h^2+8e^{i\delta}fh+4f^2)\lambda^4&(f+e^{-i\delta}h)\lambda^2-\frac{1}{2}e^{-i\delta}h\lambda^4\\
f\lambda^3&-(f+e^{i\delta}h)\lambda^2+\frac{1}{2}f\lambda^4&1-\frac{1}{2}(f^2+2e^{-i\delta}fh+h^2)\lambda^4
\end{array}\right).\nonumber
\end{eqnarray}
With this expression we can get the sides and the angles in Fig.~\ref{fig1} as
\begin{eqnarray}
AB&=&f\lambda^3;\nonumber\\
AB^{\prime}&=&f\lambda^3-\frac{1}{2}f\lambda^5;\nonumber\\
AC&=&h\lambda^3-\frac{1}{2}h\lambda^5;\nonumber\\
AC^{\prime}&=&h\lambda^3;\nonumber\\
BC&=&\lambda^3\kappa^{\frac{1}{2}}-\frac{1}{2}\lambda^5(h^2+fh)\kappa^{-\frac{1}{2}};\nonumber\\
B^{\prime}C^{\prime}&=&\lambda^3\kappa^{\frac{1}{2}}-\frac{1}{2}\lambda^5(f^2+fh)\kappa^{-\frac{1}{2}};\nonumber\\
\phi_1&=&\arctan{\frac{h\sin{\delta}}{f+h\sin{\delta}}}-\frac{fh\lambda^2\sin{\delta}}{2\kappa}+fh\lambda^4\sin{\delta}(1-\frac{fh\cos{\delta}+h^2}{4\kappa^2});\nonumber\\
\phi_1^{\prime}&=&\arctan{\frac{h\sin{\delta}}{f+h\sin{\delta}}}+\frac{fh\lambda^2\sin{\delta}}{2\kappa}+\frac{fh\lambda^4\sin{\delta}(f^2+fh\cos{\delta})}{4\kappa^2};\nonumber\\
\phi_3&=&\arctan{\frac{f\sin{\delta}}{h+f\sin{\delta}}}+\frac{fh\lambda^2\sin{\delta}}{2\kappa}+\frac{fh\lambda^4\sin{\delta}(h^2+fh\cos{\delta})}{4\kappa^2};\nonumber\\
\phi_3^{\prime}&=&\arctan{\frac{f\sin{\delta}}{h+f\sin{\delta}}}-\frac{fh\lambda^2\sin{\delta}}{2\kappa}+fh\lambda^4\sin{\delta}(1-\frac{fh\cos{\delta}+f^2}{4\kappa^2});\nonumber\\
\phi_2&=&\phi_2^{\prime}=\pi-\delta-fh\lambda^4\sin{\delta},\nonumber
\end{eqnarray}
in which $\kappa=f^2+2fh\cos{\delta}+h^2$. In the expressions for
angles (except the common inner angle $\phi_2$), the terms
proportional to $\lambda^2$ come from the fraction of the high order
terms of the elements since the definition of the angles is, for
example
$\phi_1={\mathrm{Arg}}[-\frac{V_{cd}V_{cb}^*}{V_{td}V_{tb}^*}]$,
thus we do not have this kind of corrections when we only consider
the leading order.


\begin{thebibliography}{99}

\bibitem{cabibbo}
N.~Cabibbo, Phys. Rev. Lett. 10 (1963) 531.

\bibitem{km}
M.~Kobayashi and T.~Maskawa, Prog. Theor. Phys. 49 (1973) 652.

\bibitem{ck}
L.L.~Chau and W.Y.~Keung, Phys. Rev. Lett. 53 (1984) 1802.

\bibitem{pdg}
Particle Data Group, C.~Amsler {\it et al.}, Phys. Lett. B 667
(2008) 1.

\bibitem{he}
P.H.~Frampton and X.-G.~He, Phys. Lett. B 688 (2010) 67.

\bibitem{lishiwen}
S.-W.~Li and B.-Q.~Ma, Phys. Lett. B 691 (2010) 37.

\bibitem{ut}
J.D.~Bjorken, Nucl. Phys. B (Proc. Suppl.) 11 (1989) 325.

\bibitem{boom}
P.H.~Frampton and X.-G.~He, Phys. Rev. D 82 (2010) 017301;\\
A.~Dueck, S.T.~Petcov and W.~Rodejohann, Phys. Rev. D 82 (2010) 013005.

\bibitem{maximalviolation}
D.~Hochberg and R.~Sachs, Phys. Rev. D 27 (1983) 606;\\
B.~Stech, Phys. Lett. B 130 (1983) 189;\\
I.~Dunietz, O.W. Greenberg and D.-d.~Wu, Phys. Rev. Lett. 55 (1985) 2935;\\
Y.~Koide, M.~Nakahara and C.~Hamzaoui, US-91-04, SU-LA-91-03.

\bibitem{koide}
Y.~Koide, Phys. Lett. B 607 (2005) 123;\\
Y.~Koide, Phys. Rev. D 73 (2006) 073002.

\bibitem{xing}
H.~Fritzsch and Z.-z.~Xing, Phys. Lett. B 413 (1997) 396.





\bibitem{triminimalneutrino}
S.~Pakvasa, W.~Rodejohann and T.J.~Weiler, Phy. Rev. Lett. 100
(2008) 111801.

\bibitem{triminimalquark}
X.-G.~He, S.-W.~Li and B.-Q.~Ma, Phys.\ Rev.\  D {\bf 78} (2008)
111301(R).

\bibitem{triminimalquark2}
X.-G.~He, S.-W.~Li and B.-Q.~Ma, Phys. Rev. D 79 (2009) 073001.

\bibitem{wolf}
L.~Wolfenstein, Phys. Rev. Lett. 51 (1983) 1945.



\bibitem{jarlskog}
C.~Jarlskog, Phys. Rev. Lett. 55 (1985) 1039; Z. Phys. C 29 (1985) 491.

\bibitem{Wu:1985ea}
  D.-d.~Wu,
  Phys.\ Rev.\  D {\bf 33} (1986) 860.

\bibitem{maximalcp}
A.~Hocker, H.~Lacker, S.~Laplace and F.Le~Diberder, Eur. Phy. J. C 21 (2001) 225;\\
J.~Charles, {\it et al.}, Eur. Phy. J. C 41 (2005) 1;\\
M.~Bona, {\it et al.}, JHEP 07 (2005) 028.


\end{thebibliography}
\end{document}